\documentclass[prd,twocolumn,nofootinbib]{revtex4}

\usepackage{amsmath}
\usepackage{graphicx}
\newcommand{\avg}[1]{\left\langle{#1}\right\rangle}
\def\abar{\bar\alpha}
\def\erfc{{\rm {erfc}}}
\def\xb{\tilde x}
\def\yb{\tilde y}
\def\creation{\lambda}

\begin{document}
\title{Consequences of strong fluctuations on high-energy QCD evolution}

\author{C. Marquet}
\email{marquet@spht.saclay.cea.fr}
\author{R. Peschanski}
\email{pesch@spht.saclay.cea.fr}
\author{G. Soyez\footnote{on leave from the PTF group of the University of Li\`ege.}}
\email{gsoyez@spht.saclay.cea.fr}
\affiliation{Service de physique th{\'e}orique, CEA/Saclay,
  91191 Gif-sur-Yvette cedex, France\footnote{%
URA 2306, unit\'e de recherche associ\'ee au CNRS.}}

\begin{abstract}
We investigate the behaviour of the QCD evolution towards high-energy, in the 
diffusive approximation, in the limit where the fluctuation contribution is large. 
Our solution for the equivalent stochastic Fisher equation predicts the 
amplitude as well as the whole set of correlators in the strong noise limit. 
The speed of the front and the diffusion coefficient are obtained. 
We analyse the consequences on high-energy evolution in QCD.
\end{abstract}

\maketitle


1. The quest for the perturbative high-energy limit of QCD has been the subject of many efforts. It is now well accepted that, due to the strong rise of the amplitude predicted by the linear Balitsky-Fadin-Kuraev-Lipatov (BFKL) equation \cite{bfkl}, one has to include saturation effects in order to describe high parton densities and recover unitarity. In the large-$N_c$ limit and in the mean-field approximation, we are led to consider the Balitsky-Kovchegov (BK) equation \cite{bk}.
It has the nice property \cite{mp} to be mapped, in the diffusive approximation, onto the Fisher-Kolmogorov-Petrovsky-Piscounov (F-KPP) equation \cite{fkpp} which has been widely studied in statistical physics and is known to admit traveling waves as asymptotic solutions, translating into geometric scaling in QCD \cite{geomsc}.


It has recently been proven \cite{fluct,it} that fluctuation effects have important consequences on the approach to saturation. Practically, the resulting evolution equation, after a coarse-graining approximation \cite{it}, takes the form of a Langevin equation. It is formally equivalent to the BK equation supplemented with a noise term
\begin{eqnarray}
\partial_Y T(L,Y) & = &\abar\chi(-\partial_L)T(L,Y) - \abar T^2(L,Y) \nonumber\\
                  & + & \abar\sqrt{\kappa\alpha_s^2 T(L,Y)} \nu(L,Y)\label{eq:langevin}
\end{eqnarray}
where $T$ is the event-by-event scattering amplitude, $Y$ is the rapidity, $L=\log(k^2/k_0^2)$ with $k$ the transverse momentum and $k_0$ some arbitrary reference scale. $\chi(\gamma)=2\psi(1)-\psi(\gamma)-\psi(1-\gamma)$ is the BFKL kernel, $\kappa$ is a fudge factor and the noise $\nu(L,Y)$ satisfies $\avg{\nu}=0$ and
\begin{equation}\label{eq:noisecorel}
\avg{\nu(L,Y)\nu(L',Y')} = \frac{2}{\pi} \delta(\abar Y-\abar Y')\delta(L-L').
\end{equation}
To obtain equation \eqref{eq:langevin}, we have worked with impact-parameter-independent amplitudes, for which the original non-local, off-diagonal noise term takes \cite{it,msw} the form given by equation \eqref{eq:noisecorel}.
Physically, in addition to the BK saturation effects coming from pomeron merging in the target, one also takes into account pomeron splitting in the target. Hence, by combinations of splittings and mergings, pomeron loops are formed. The extra factor $\alpha_s^2$ in the fluctuation term indicates that it is dominant when $T\sim \alpha_s^2$ \textit{i.e.} in the dilute limit. 

In the same line that the BK equation is equivalent to the F-KPP equation in the diffusive approximation, the Langevin problem corresponds to the stochastic F-KPP (sFKPP) equation \cite{duality}. 
To be more precise, let us expand the BFKL kernel to second order around $\gamma_0$
\begin{equation}
\chi(\gamma) = \chi_0 + \chi'_0(\gamma-\gamma_0)+\frac{1}{2}\chi''_0(\gamma-\gamma_0)^2
             = A_0 + A_1\gamma + A_2\gamma^2.\label{eq:coefsai}
\end{equation}
This approximation has proven its ability to exhibit the main properties of the solutions of equation \eqref{eq:langevin} in the limit $\kappa\alpha_s^2 \ll 1$. Unless specified, we shall keep this approximation throughout this paper, leaving the general case for further studies.

If we introduce \textit{time} and \textit{space} variables as follows
\[
t = \abar Y,\quad x = L-A_1\abar Y\quad\text{and }u(x,t)=\frac{T}{A_0},
\]
equation \eqref{eq:langevin} gets mapped onto the sFKPP equation \footnote{We have introduced and extra factor $\sqrt{1-u}$ in the noise term. The effects of the noise being important in the dilute tail, this modification is not expected to change physical results. In addition, in \eqref{eq:langevin}, the fluctuation contribution is only under control in the dilute regime.}
\begin{equation}\label{eq:sfkpp}
\partial_t u = D \partial_x^2 u + \creation u(1-u) + \varepsilon\sqrt{u(1-u)}\eta(x,t),
\end{equation}
with
\begin{eqnarray*}
&&D=A_2, \quad \creation=A_0, \quad \varepsilon^2=\frac{2\kappa\alpha_s^2}{\pi A_0},\\
&&\avg{\eta(x,t)\eta(x',t')} = \delta(t-t')\delta(x-x').
\end{eqnarray*}


At present stage, most of the analytical analysis were performed in the limit where the fluctuations are asymptotically small (the correction being logarithmic \cite{brunet}, it may require a strong coupling constant as small as $\alpha_s \lesssim 10^{-10}$), in which case the relevant quantities, \textit{e.g.} the speed of the wave, can be expanded around the F-KPP solution.
In this analysis, the main effects of the noise in the sFKPP equation are, first, to lead to a decrease of the speed of the traveling front and, second, to produce a diffusion of the fronts for each realisation of the noise resulting in violations of geometric scaling for the average amplitude.

The numerical studies performed so far show that these effects (decrease of the speed and diffusion of the events) are amplified when the fluctuation coefficient becomes more important. There has been large efforts made to improve the analytical understanding for arbitrary values of the noise strength but many approaches appear to be model-dependent \cite{panja}.

In this paper, we consider the problem of fluctuations through a complementary approach, namely the limit of a strong noise. This limit is tractable with the help of a \textit{duality} property of the sFKPP equation \cite{duality}. The strong-noise limit then gets related to a coalescence process which can be solved exactly \cite{coal}. 

Using these tools from statistical physics, we are able to compute the event-averaged amplitude as well as the higher-order correlators and obtain predictions for the speed of the wavefront as well as for the diffusion coefficient in the limit of strong fluctuations. This knowledge of the strong-noise limit, together with the weak-noise results, can help in further analytical understanding of the QCD evolution in the presence of fluctuations.


2. Let us now summarise the tools from statistical physics we need for our studies. Our starting point is the duality relation between the sFKPP Langevin equation and the reaction-diffusion process. This duality will allow us to relate the strong noise problem to the coalescence problem which we shall solve.

We consider on the one hand amplitudes evolving according to the sFKPP equation \eqref{eq:sfkpp} and, on the other hand, the reaction-diffusion process of a one-dimensional population $A$ on a lattice of spacing $h$: at each site, one can have particle creation or recombination, and particles can diffuse to a neighbouring site
\begin{equation}\label{eq:partic}
A_i \stackrel{\creation}{\to} A_i+A_i, \quad
A_i+A_i \stackrel{\varepsilon^2/h}{\longrightarrow} A_i \quad\text{and }\quad
A_i \stackrel{D/h^2}{\longrightarrow} A_{i\pm 1}
\end{equation}
where $A_i$ designs a particle at lattice site $i$.

One shows \cite{duality} that the particle system and the amplitude in the sFKPP equation are related through a duality relation which, in the continuum limit $h\to 0$, can be written
\begin{equation}\label{eq:duality}
\avg{\prod_x\left[1-u(x,t)\right]^{N(x,0)}} = \avg{\prod_x\left[1-u(x,0)\right]^{N(x,t)}},
\end{equation}
where $N(x,t)$ is the particle density in the reaction-diffusion system.
Physically, this duality equation means that, if one wants to obtain the scattering amplitude at rapidity $t=\abar Y$, one can either keep the target fixed and evolve the projectile wavefunction considered as a particle system, or fix the projectile and consider evolution of scattering amplitudes off the target. One knows \cite{ddd,ist} that splitting in the projectile leads to linear growth and saturation in the target while merging in the projectile corresponds to fluctuations in the target.


Therefore, to obtain information on the evolution of the average amplitudes for the sFKPP equation, we shall study the dual particle system and then use relation \eqref{eq:duality}. The limit we are interested in is the strong noise limit (large $\varepsilon$). This corresponds to heavy saturation in the particle system (projectile), \textit{i.e.} two particles at the same lattice site automatically recombine into a single one. This limit is often referred to as the \textit{diffusion-controlled limit}. When $\creation$ is rescaled to give a constant, small, ratio $\creation/\varepsilon^2$, we can alternatively study the \textit{coalescence model}. In this model, one can have at most one particle per lattice site. One particle can diffuse to the neighbouring site at rate $D/h^2$ or give birth to a new one in a neighbouring site at rate $\omega/h$ (with $\omega=2D\creation/\varepsilon^2$ as we shall see later).

This system has been studied \cite{coal} and is fully solvable using the method of \textit{inter-particle probability distribution function}. The main idea is to introduce $E(x, y; t)$ as the probability that sites between $x$ and $y\ge x$ included are empty at time $t$. One obtains that, due to diffusion and creation, $E$ satisfies the following differential equation
\begin{equation}\label{eq:sysevol}
\partial_t E = \left\{ D\left(\partial_x^2+\partial_y^2\right) 
                    + \omega\left(\partial_y-\partial_x\right) \right\} E.
\end{equation}
with the boundary condition $\lim_{y\to x} E(x,y;t) = 1$.

The particle density can be obtained from the derivative of $E$:
\[
N(x,t) = \left.\partial_yE(x,y;t)\right|_{y\to x}.
\]
Remarkably enough, the evolution equation \eqref{eq:sysevol} is linear.  It can be solved exactly \cite{coal} and, for a given initial condition $E(x,y;t)$, introducing the dimensionless variables
\[
\xi = \frac{\omega}{D}(x+y),\quad \zeta=\frac{\omega}{D}(y-x)\quad\text{ and } \tau = \frac{8\omega^2}{D}t,
\]
one finds
\begin{eqnarray}\label{eq:syssol}
E(x,y;t) & = & e^{-\zeta} + e^{-\tau}\int_{-\infty}^\infty d\xi'\int_0^\infty d\zeta'\\
&  & \,G(\xi,\xi',\zeta,\zeta';\tau)\left[E(\xi',\zeta';\tau)-e^{-\zeta'}\right],\nonumber
\end{eqnarray}
where the Green function $G(\xi,\xi',\zeta,\zeta';\tau)$ is given by
\[
\frac{1}{\pi\tau}e^{-(\xi-\xi')^2/\tau} e^{-(\zeta-\zeta')/2}
\left[e^{-(\zeta-\zeta')^2/\tau}-e^{-(\zeta+\zeta')^2/\tau}\right].
\]

Before considering the solution of this system in the context of the duality relation, let us give the relation between the parameters $\creation$ and $\varepsilon^2$ of the initial system with $\omega$ and $D$ in the coalescence model. The trick is to require that both systems have the same equilibrium density. For the coalescence model, one notice that $\exp(-\frac{\omega}{D}(y-x))$ is a time-independent solution leading to a particle density $N_{\text{eq}}=\omega/D$. In the case of process \eqref{eq:partic}, at equilibrium, diffusion does not play any role and we have to find equilibrium at each site between creation end annihilation. This is achieved when $N_{\text{eq}} = 2\creation/\varepsilon^2$. Hence, one has $\omega=2D\creation/\varepsilon^2$.



3. Let us now put together the results from duality and coalescence and derive the sFKPP solution. 

By carefully choosing the initial condition, equation \eqref{eq:duality} simplifies. If one starts with one particle at position $x$ \textit{i.e.} $N(\xb,0) = \delta(\xb-x)$, then the l.h.s. of \eqref{eq:duality} becomes simply $1-\avg{u(x,t)}$. By starting with $k$ particles at position $x_1,\dots,x_k$, one similarly obtains $\avg{[1-u(x_1,t)]\dots[1-u(x_k,t)]}$. 

In addition, let us start with a step function for the amplitude $u(x,0)=\theta(-x)$. Then $\avg{\prod_x\left[1-u(x,0)\right]^{N(x,t)}}$ is the probability that, in the particle process, all sites $x\le 0$ are empty. For the case of the strong noise \textit{i.e.} when the particle system is the coalescence model, this probability is directly obtained in terms of the density $E$ and the duality relation becomes
\[
\avg{u(x,t)}=1-E_x(-\infty,0;t),
\]
with the initial condition
\[
E_x(\xb,\yb;0)=1-\theta(x-\xb)\theta(\yb-x).
\]

Inserting this initial condition inside the general solution \eqref{eq:syssol}, we get after a bit of algebra
\begin{widetext}
\begin{eqnarray*}
E_x(\xb,\yb;t) & = & \frac{1}{2}
\left\{
\erfc\left(\frac{\yb\!-\!\xb\!+\!2\omega t}{\sqrt{8Dt}}\right)
\!+\!\erfc\left(\frac{\xb\!-\!\yb\!-\!2\omega t}{\sqrt{8Dt}}\right)
\!-\!\erfc\left(\frac{x\!-\!\yb\!-\!\omega t}{2\sqrt{Dt}}\right)
\left[1\!-\!\frac{1}{2}\erfc\left(\frac{x\!-\!\xb\!+\!\omega t}{2\sqrt{Dt}}\right) \right]
\right\}\\
& + & \frac{1}{2}e^{-\frac{\omega}{D}(\yb-\xb)}
\left\{
2\!-\!\erfc\left(\frac{\yb\!-\!\xb\!-\!2\omega t}{\sqrt{8Dt}}\right)
\!-\!\erfc\left(\frac{\xb\!-\!\yb\!+\!2\omega t}{\sqrt{8Dt}}\right)
\!+\!\erfc\left(\frac{x\!-\!\xb\!-\!\omega t}{2\sqrt{Dt}}\right)
\left[1\!-\!\frac{1}{2}\erfc\left(\frac{x\!-\!\yb\!+\!\omega t}{2\sqrt{Dt}}\right) \right]
\right\}
\end{eqnarray*}
\end{widetext}
where $\erfc(x)$ is the complementary error function.
The limit $\xb\to -\infty$, $\yb\to 0$ in this expression gives
\begin{eqnarray}\label{eq:sol}
\avg{u(x,t)} & = & \frac{1}{2}\erfc\left(\frac{x-\omega t}{2\sqrt{Dt}}\right)\\
& = &\frac{1}{2\sqrt{D\pi t}}\int_{-\infty}^\infty d\xb\,\theta(\xb-x)
                         e^{-\frac{(\xb-\omega t)^2}{4Dt}}.\nonumber
\end{eqnarray}

This results calls for comments. First, it corresponds to a wave traveling at an average speed 
\[
\omega=\frac{2D\creation}{\varepsilon^2}.
\]
This result confirms the decrease of the velocity but contrasts with the speed obtained in the weak noise limit by perturbative analysis around the F-KPP speed $\simeq 2\sqrt{D\lambda}-\pi^2\sqrt{D\lambda} |\log(\varepsilon)|^{-2}.$

The expression \eqref{eq:sol} shows that the amplitude could be obtained from a superposition of step functions around $x=\omega t$ with a Gaussian noise of width $\sqrt{2Dt}$. The interesting point here lies in the dispersion coefficient: in the weak-noise analysis, it behaves like $|\log(\varepsilon)|^{-3}$. We have proven that this rise goes to $2D$ when the noise becomes strong.

In addition, one can probe the correlators of the amplitude by starting with an initial condition with particles at positions $x_{\text{min}}=x_1 < \dots < x_k=x_{\text{max}}$:
\[
E(x,y;0) = 1-\sum_{i=1}^k \theta(x-x_{i-1})\theta(x_i-x)\theta(y-x_i),
\]
with, formally, $x_0=-\infty$.
Following the same lines as above, one gets
\[
\avg{[1-u(x_1,t)]\dots[1-u(x_k,t)]} = 1-\frac{1}{2}\erfc\left(\frac{x_{\text{min}}-\omega t}{2\sqrt{Dt}}\right).
\]
We need to use this relation to obtain the correlations of $u$ instead of $1-u$. This is obtained as follows
\begin{eqnarray*}
\lefteqn{\avg{u(x_1,t)\dots u(x_k,t)}}\\
& = & \avg{\{1-[1-u(x_1,t)]\}\dots \{1-[1-u(x_k,t)]\}}\\
& = & \sum_{A\subseteq \{1,\dots,k\}} (-)^{\sharp A}\avg{\prod_{i\in A}[1-u(x_i,t)]}\\
& = & \sum_{A\subseteq \{1,\dots,k\}} (-)^{\sharp A}\avg{1-u(x_{\min(A)},t)}
\end{eqnarray*}
where $\sharp A$ is the cardinal of the set $A$ and $\min(A)$ is its minimum. The sum can be reordered to give
\begin{eqnarray*}
\lefteqn{\avg{u(x_1,t)\dots u(x_k,t)}}\\
& = & 1-\sum_{j=1}^k \avg{1-u(x_j,t)}\sum_{A\subseteq \{j+1,\dots,k\}} (-)^{\sharp A}\\
& = & 1-\sum_{j=1}^k [1-\avg{u(x_j,t)}]\sum_{n=0}^{k-j} (-)^n\binom{k-j}{n},
\end{eqnarray*}
In the last expression, only the term with $j=k$ survives hence (with $x_{\text{max}}=x_k$)
\[ 
\avg{u(x_1,t)\dots u(x_k,t)} = \frac{1}{2}\erfc\left(\frac{x_k-\omega t}{2\sqrt{Dt}}\right)
                             = \avg{u(x_k,t)}.
\] 
This simple result can again be seen as a superposition of step functions with a Gaussian dispersion. $u(x_1)\dots u(x_k)$ is nonzero provided $u$ does not vanishes at the position with largest coordinate ($x_k$). Hence, the whole dynamics is the same as if only one particle were lying at position $x_k$ in the initial condition \footnote{Of course, the same argument holds for the product of $1-u(x_i)$ and the fronts around $x_1$.}.

\begin{figure}
\includegraphics[scale=0.9]{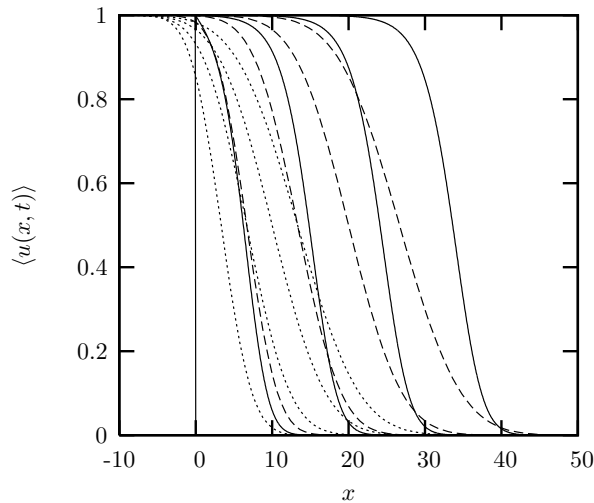}
\caption{Comparison between the F-KPP and sFKPP wavefronts at $t=0,5,10,15$ and $20$. Solid curve: numerical simulation of the F-KPP equation. Dashed curve: analytic result for $\varepsilon^2=1.5$. Dotted curve: analytic result for $\varepsilon^2=3$.}\label{fig:front}
\end{figure}

In order to illustrate the effect of the noise, we have plotted in figure \ref{fig:front} the time evolution of the wavefront $\avg{u(x,t)}$ for numerical simulations of the F-KPP equation and for our solution \eqref{eq:sol} ($\varepsilon^2=1.5$ and $3$). In each case, the initial condition is a step function. We clearly see that the effect of the strong noise is to slow down considerably the front and add significant distortion.

4. Coming back to QCD variables, we find (assuming without loss of generality $L_1\le\dots\le L_n$)
\begin{eqnarray}\label{eq:front}
\avg{T(L_1,Y)\dots T(L_k,Y)}
 & = & A_0^{k-1}\avg{T(L_k,Y)}\\
 & = & \frac{A_0^k}{2}\erfc\left(\frac{L_k-c\abar Y}{\sqrt{2D_{\text{diff}}\abar Y}}\right)
\nonumber
\end{eqnarray}
with the speed of the traveling front and the diffusion coefficient
\begin{equation}\label{eq:res}
c = A_1+\frac{\pi A_2A_0^2}{\kappa\alpha_s^2}\qquad\text{and } D_{\text{diff}} = 2A_2.
\end{equation}

Let us discuss the physical interpretation of these results. We start by equation \eqref{eq:front}. It is remarkable that the strong-fluctuation limit gives, as an analytic solution, the asymptotic result inferred in previous studies \cite{it,himst}. It proves the universal feature that high-energy scattering amplitudes are described by a superposition of Heavyside functions with Gaussian dispersion. We confirm analytically that the correlators are driven by the amplitude of the largest momentum in the process. This is the main result of this paper.

Equation \eqref{eq:res} relates the parameters of the amplitude {\it e.g.} the average speed $c$ and the diffusion constant $D_{\text{diff}}$ to the parameters $A_0$, $A_1$ and $A_2$ obtained from the expansion \eqref{eq:coefsai} of the BFKL kernel. If one performs this expansion choosing $\gamma_0=\scriptstyle{\frac{1}{2}}$ or $\gamma_0=\gamma_c \approx 0.6275$, as used in the weak-noise limit, the value of $A_1$ turns out to be negative. This would lead to a negative speed which seems unphysical in QCD. The way out of this inconsistency is to insert the solution \eqref{eq:front} directly into the exact evolution for $\avg{T(L,Y)}$. It has been shown \cite{himst} that it results, as expected, into a vanishing speed. The determination of $c$ through the evolution equation would depend on the corrections to the error function, which disappear in the strong-noise limit. 

Let us sketch a heuristic argument indicating that a physically meaningful behaviour of these parameters can be obtained in the limit $\gamma_0$ small. Indeed, this is suggested by the fact that, in the strong-noise limit, each front is approaching a Heavyside function \eqref{eq:sol} which suggests to perform the kernel expansion near $\gamma_0=0$. Considering $\gamma_0$ small, one has $A_0=3\gamma_0^{-1}$, $A_1=-3\gamma_0^{-2}$  and $A_2=\gamma_0^{-3}$, leading to
\[
c = \frac{3}{\gamma_0^2}\left(\frac{3\pi}{\gamma_0^3\kappa\alpha_s^2}-1\right).
\]
If one requires $c\to 0$, one has to choose
\[
\gamma_0 \approx \left(\frac{3\pi}{\kappa\alpha_s^2}\right)^{1/3}\left\{1-o\left[\left(\frac{3\pi}{\kappa\alpha_s^2}\right)^{\frac{2}{3}}\right]\right\}
\]
where $o(x)$ denotes a function falling to zero faster than $x$. The physical parameters, in the strong-noise limit, are then
\begin{equation}\label{eq:newres}
c\to 0 \qquad\text{and } D_{\text{diff}} = \frac{2\kappa\alpha_s^2}{3\pi}.
\end{equation}
It is interesting to notice that, if we choose $\gamma_0$ to satisfy the only requirement that $c\to 0$, the diffusion coefficient is entirely determined, independently of the way this limit is achieved.

On a more general ground, it is worth noting that the strong-noise development should not be considered as a strong coupling expansion since the initial equation \eqref{eq:langevin} is derived in a perturbative framework. Indeed, the genuine expansion parameter in the strong noise limit is $\kappa \alpha_s^2$ where $\kappa$ could reach high values in the physical domain of interest \cite{dt}. Thus, even in the perturbative regime, the noise parameter $\kappa\alpha_s^2$ may be large. These predictions, together with the weak noise results, enclose the physical domain of $\kappa\alpha_s^2$. The knowledge of both these limits, together with information from numerical estimates \cite{simul}, can lead to a better understanding of the physics of fluctuations. An expansion in $1/(\kappa\alpha_s^2)$ could also lead to faster convergence than the logarithmic weak-noise expansion \cite{bdmm}.

5. Let us summarise our results. We have used the duality relation between the amplitude given by the stochastic FKPP equation and the particle densities in a reaction-diffusion process. This duality is physically similar to the projectile-target duality noticed recently in high-energy QCD when both saturation and fluctuation effects are taken into account. The saturation in the target is related to splitting in the projectile while fluctuations are mapped to recombination in the particle system.

In the case of large fluctuations \textit{i.e.} strong noise in the sFKPP equation, the corresponding particle system can be described as a coalescence problem. This process can be solved exactly using the inter-particle probability distribution function. We use this to compute the average value of the amplitude as well as the correlators. 

The main result of our analysis is the analytic derivation of the average scattering amplitude as a universal error function \eqref{eq:front} which also determines higher-order correlators. This picture proposed in previous studies is thus confirmed and shows that the results obtained in the limit of strong fluctutions possess a physical meaning.

The fact that the correlators display the same behaviour as the amplitude itself is physically interesting. This is obtained through a superposition of event-by-event amplitudes which are 0 or 1. The dominant contribution to the scattering process comes when all individual scattering are 1. This picture of a black and white target gives rise to new scaling laws for Deep Inelastic Scattering \cite{himst}.

Since the physically acceptable values of the strong coupling seem to lie in between the strong and the weak noise limits, a knowledge of both approaches is useful.

As an outlook, it would be interesting to extend our formalism beyond the diffusive approximation and/or to modified evolution kernels \cite{param}. This may allow a better determination of the speed and dispersion coefficient, given by \eqref{eq:newres} in the diffusive approximation. Also, a perturbative approach starting from the strong noise limit could prove useful. These questions certainly deserve further studies.

\begin{acknowledgments}
G.S. is funded by the National Funds for Scientific Research (Belgium).
\end{acknowledgments}

\end{document}